\documentclass[5p]{elsarticle}

\usepackage{hyperref}
\usepackage{graphicx}
\usepackage{xcolor}

\journal{Astronomy \& Computing}

\biboptions{authoryear}

\begin{document}

\begin{frontmatter}

\title{SAMP, the Simple Application Messaging Protocol:
       Letting applications talk to each other}

\author[bristol]{M.~B.~Taylor}
\ead{m.b.taylor@bristol.ac.uk}
\author[cds]{T.~Boch}
\author[roe]{J.~Taylor\fnref{google}}

\address[bristol]{H.~H.~Wills Physics Laboratory, University of Bristol, UK}
\address[cds]{CDS, Observatoire Astronomique de Strasbourg, France}
\address[roe]{Institute for Astronomy, Royal Observatory, Blackford Hill,
              University of Edinburgh, UK}
\fntext[google]{Present address: Google, USA}

\begin{abstract}
SAMP, the Simple Application Messaging Protocol, is a
hub-based communication standard for the exchange of data and control
between participating client applications.
It has been developed within the context of the Virtual Observatory
with the aim of enabling specialised data analysis tools to cooperate
as a loosely integrated suite, and is now in use by many and varied
desktop and web-based applications dealing with astronomical data.
This paper reviews the requirements and design principles that led to
SAMP's specification, provides a high-level description of the protocol,
and discusses some of its common and possible future usage patterns,
with particular attention to those factors that have aided its success
in practice.
\end{abstract}


\end{frontmatter}


\newcommand{\kdot}{.\linebreak[0]}

\section{Introduction}

Astronomical research requires complex and flexible manipulation
and processing of various different types of data.
Images, spectra, catalogues, time series, coverage maps and other data
types need their own special handling,
typically provided by specialist tools.
Data sets of different types meanwhile are usually related
in various ways arising from their physical origin,
for instance
catalogues are often derived from images and best understood in
conjunction with them, and
spectra and time series usually originate from specific sky positions
or regions which may be represented on images and described by
catalogue entries.
To extract scientific meaning from the data it is usually necessary
to exploit these linkages between data items as well as the
internal structure of each.

The working astronomer therefore uses a selection
of different software components,
each specialising in a particular type of data or manipulation,
for different data sets and different tasks,
and has to integrate these together in a way that takes account of
the relationships of the data items under consideration.

For batch or pipeline-type processing the required tool integration
is usually, in terms of data flow, fairly straightforward:
the output of one step can be fed to the input of the next as a file,
stream of bytes, or some kind of parameter list,
often under the control of a script of some kind.

During the exploratory or interactive phase of data analysis however,
this traditional model of tool integration is less satisfactory.
Within a given GUI analysis application
it is usual to interact with the data using
mouse and keyboard gestures to perform actions like selection or
navigation with instant visual feedback, in many cases with some
kind of internal linkage between different data views.
But communicating such actions or their results between different
tools tends to be much more cumbersome.
A way can often be found to reflect a result generated by one tool
in the state of another, for instance by reading sky coordinates
reported by one tool and typing or pasting them into another,
or saving an intermediate result from one tool to temporary storage
and reloading it into another, but it can be fiddly and tedious,
especially if similar actions are required repeatedly.
This lack of convenience is more than just an annoyance, it can
interrupt the flow of the data exploration, reduce the parameter
space able to be investigated, and effectively stifle discovery
of relationships present in the data.

From this point of view, a single monolithic astronomical data analysis
user application providing the best available facilities for
interactive presentation, manipulation and analysis of all kinds of
astronomical data and their interrelationships seems an attractive prospect.
In reality of course, no such one-stop analysis tool exists.
The obvious practical difficulties aside, it is not even clear
that deviating so far from the Unix philosophy of
``Make each program do one thing well'' \citep{mcilroy1978}
would be desirable.

These considerations have driven the development of a framework
for communication between independently-developed software items,
written in different languages and running in different processes.
Such applications can thus be made to appear to the user
as a loosely integrated suite of cooperating tools,
providing facilities such as data exchange, linked views and
peer-to-peer or client-server remote control.
Although communication between interactive desktop tools was the
original stimulus for what is now SAMP, the framework is flexible
enough to support other usage patterns as well.

Two previous papers on SAMP have been presented in the ADASS conference
series:
\citet{adassxxi_paper} briefly outlines the architecture and explains
the Web Profile, and \citet{adassxxii_bof} lists some existing
client libraries.
The current paper discusses the protocol,
its communication model, and its current usage
in sufficient detail to understand the design decisions taken
and their consequences,
particularly from the point of view of the usage scenario
outlined above.
Section \ref{sec:context} traces the evolution of SAMP from
its predecessor PLASTIC alongside a comparison with some
alternative messaging systems,
section \ref{sec:design} outlines some high-level design principles,
section \ref{sec:protocol} presents a description of the protocol itself
along with some of the thinking behind it,
section \ref{sec:usage} considers its use in practice, and
section \ref{sec:conclusion} concludes by reviewing the current status
and possible future directions for SAMP,
as well as the factors that have encouraged its uptake.
For the complete and definitive details of the protocol,
the reader is referred to the standard document itself \citep{samp_std}.

\section{Context} \label{sec:context}

\subsection{History}

In the context of the emerging Virtual Observatory in the mid-2000s,
the benefit of connecting client-side tools to improve productivity
when working with multiple data types became apparent.
In fact this problem was not specific to the VO, 
but the ease with which multiple related data products
could be acquired using VO technologies,
themselves sometimes requiring the use of separate tools for
data discovery and acquisition,
amplified the benefits that such tool integration could deliver.
Additionally, the new shared funding and communications channels
between institutionally and geographically separated software developers
that arose from various VO initiatives
proved important in practice
as a platform for experimentation and agreement in this area.

The external scripting capabilities of tools such as
Aladin \citep{2000A&AS..143...33B},
SPLAT-VO \citep{ACVOsplat}
and SAOImage ds9 \citep{2003ASPC..295..489J}
already provided the option of tightly coupled master/slave
control between pairs of applications, but did not lend themselves
to the kind of cooperative interaction envisaged.
The developers of Aladin experimented with Java interfaces designed
for two-way communications; these delivered some limited integration,
but were restricted to applications operating within the same Java
virtual machine.
Meanwhile the Astro Runtime \citep{2007ASPC..376..571W} developed by
AstroGrid was providing to desktop tools a simplified fa\c{c}ade
for a range of VO services using their choice of communication
technology (XML-RPC, REST, Java RMI or JVM call).

From this background, in 2005 discussions between developers of
the AstroGrid, Aladin,
VisIVO \citep{2007PASP..119..898C} and
TOPCAT \citep{2005ASPC..347...29T} software
in the context of the Euro-VO
framework and the SC4DEVO workshop series led to the development
of a new communication protocol PLASTIC:
the PLatform for AStronomical Tool InterConnection
\citep{2007ASPC..376..511T,plastic_note}.

PLASTIC built on the implementation and technology choices
present in the Astro Runtime to provide the interaction
capabilities required by the participating teams,
and prototyped many features that were later inherited by SAMP,
including a central hub, publish-subscribe messaging, use of XML-RPC,
loosely-defined message semantics, and a pragmatic approach
to providing ``good-enough'' communications.
It proved popular with developers and users, 
and was incorporated into a dozen or so desktop applications,
which could thereby be used together effectively
in productive and sometimes novel ways.

Interest in PLASTIC was however largely confined to Europe.
Efforts to gain IVOA endorsement and expand the pool
of applications that could communicate in this way
led after some discussion to the drafting by European and US authors
of a successor standard, the Simple Application Messaging Protocol,
which was accepted as an IVOA Recommendation in 2009 (SAMP version 1.11).
This standard was intentionally similar in many respects to PLASTIC,
in order to avoid disrupting patterns of successful
cooperation already in use,
but the opportunity was taken to amend some decisions that experience had
shown to be sub-optimal, and to expand its scope to accommodate
other possible usage patterns.
Changes made on the basis of lessons learned from PLASTIC included
a simplification of the type system,
complete language independence
(though PLASTIC could be used from any language,
certain parts of the protocol were defined with reference to Java),
simplification of message targetting,
improved security arrangements
(security is still rudimentary in SAMP, but opportunities for
trivial client spoofing were removed),
modification of message names
(now both human-readable and wildcard-able rather than opaque URIs),
definition of all message parameters and return values as
key-value pairs rather than ordered lists,
use of fundamentally asynchronous messaging for robustness,
restriction rather than proliferation of transport mechanisms,
improved error reporting,
and better extensibility.

Also new in SAMP was the notion of a Profile to provide formal separation
between the abstract messaging model and the transport layer.
One reason for its introduction was to enable the possible future
use of the protocol for messaging in less ``PLASTIC-like'' contexts.
At the time, requirements for improved performance or security were
envisaged; to date extensions in those directions have not been
explored, but the Profile mechanism has paid off by supporting
the later development of the Web Profile to support browser-based
clients alongside desktop ones.
The introduction of the Web Profile in SAMP version 1.3 (2012)
has been the main change to date since the initial version.

\subsection{Other Messaging Systems}

Many general- and special-purpose messaging frameworks exist.
It is beyond the scope of this paper to provide a comprehensive
review, but we provide here a brief comparison between SAMP and
a few of the alternatives.

Several generic messaging frameworks share some features with SAMP, 
for example AMQP, ZeroMQ, XPA, and D-Bus.
To our knowledge, none
satisfy SAMP's key requirements of an easily implementable
platform-neutral standard supporting straightforward messaging
between a shared community of clients in quite the way required,
though some of these systems could be used as transport layers on
which future SAMP profiles could be built, in the same way
that XML-RPC has been used in the existing profiles.
SAMP deliberately restricts some choices related to implementation
and usage to reduce the burden on client developers,
so it is perhaps not surprising that generic messaging
frameworks are not, on their own, appropriate.

One or two messaging systems however merit further mention.
WAMP, the Web Application Messaging Protocol\footnote{\url{http://wamp.ws/}}, 
bears some striking architectural similarities to SAMP including 
a combination of RPC and publish-subscribe messaging mediated by 
a central component known in WAMP terminology as the Router.
However, it does not address the issue of router discovery,
so there is no prescribed way for clients to initiate communication.

The Intent mechanism for inter-process communication that forms part of
the Android operating system, while its target environment
clearly differs from that of SAMP,
shares with it some characteristics in terms of design and usage.
For instance message semantics may be defined in a way which
is either app-specific (``explicit'') or vague (``implicit''),
in the latter case resulting in a
user choice at runtime between candidate receiving apps.
Furthermore, bulk data transfer is achieved
using URIs to refer to an external data source
rather than conveying the payload within the message.

An example of a domain-specific messaging framework is
the Systems Biology Workbench \citep{sbw},
which is close in spirit to SAMP, enabling platform-independent 
remote method invocation between components (known as ``modules'') 
in the field of systems biology, based around the SBML data format.
One point of difference is that
the SBW infrastructure itself orchestrates the loading of
modules to provide the required functionality
(the Android OS does something similar with Intents);
in SAMP this choice of components is left to the user,
though components like AppLauncher
\citep{2012ASPC..461..379L} are available to
layer automatic client startup on top of the basic protocol where desired.

\section{Design for Interoperability} \label{sec:design}

The overriding objective for the design of SAMP
has been to foster interoperability in practice.
This requires not just a messaging system with sufficient communication
capabilities, but also one which developers of popular analysis tools,
and ideally private scripts as well,
are actually willing and able to integrate into their software.
To achieve this, a number of principles
have been followed.

In the first place, it is as far as possible platform independent.
The definition of the protocol is not dependent on or biassed
towards use of particular implementation languages or operating systems.

Second, ease of adoption.
Application authors have found basic use of 
SAMP to require little implementation effort.
In practice, availability of easily deployable SAMP client libraries
developed within the SAMP community
for a number of implementation languages have been an important
factor in this.
However the communications are, by design, simple enough that
basic SAMP use is not hard to achieve given only an XML parser
and HTTP access capabilities.
Ease of use by end users is equally important,
so that those running analysis tools
can benefit from the integration capabilities that SAMP provides
without needing to perform expert configuration
(ideally, any explicit setup at all) or to understand the details
of the messaging system.

Third, extensibility and flexibility.
Building into the system the capability to use it in ways
driven by the requirements of the client tools rather than
just those forseen by the standard authors increases its likely usefulness.
The mechanisms for extensibility have been particularly designed
to allow the introduction of new features without affecting existing ones,
with the aim of reducing compatibility issues.

Finally, the approach has been above all pragmatic, favouring
the straightforward over the rigorous in cases of conflict.
For instance message delivery is not guaranteed, but can be 
expected to work most of the time.
The security model will prevent casual interference,
but may be vulnerable to determined attack.
Semantics are tagged
using short readable strings on the assumption of sensible choices,
rather than URIs with guaranteed private namespaces.
Performance is easily good enough to handle exchange of short control
messages on a timescale commensurate with user actions,
but not for sustained throughput of high data volumes.
It may however be noted that most
of these items could if required be ameliorated by future
introduction of a new Profile with different transport characteristics.

These principles and their application, in some cases informed
by positive and negative lessons from the experience of PLASTIC,
might not be appropriate for all contexts
but have led to a messaging infrastructure
which ought to be easy for client developers to understand and adopt,
and which has in fact been widely taken up.

\section{Protocol Description} \label{sec:protocol}

\begin{figure}
\begin{center}
\includegraphics[width=0.9\columnwidth]{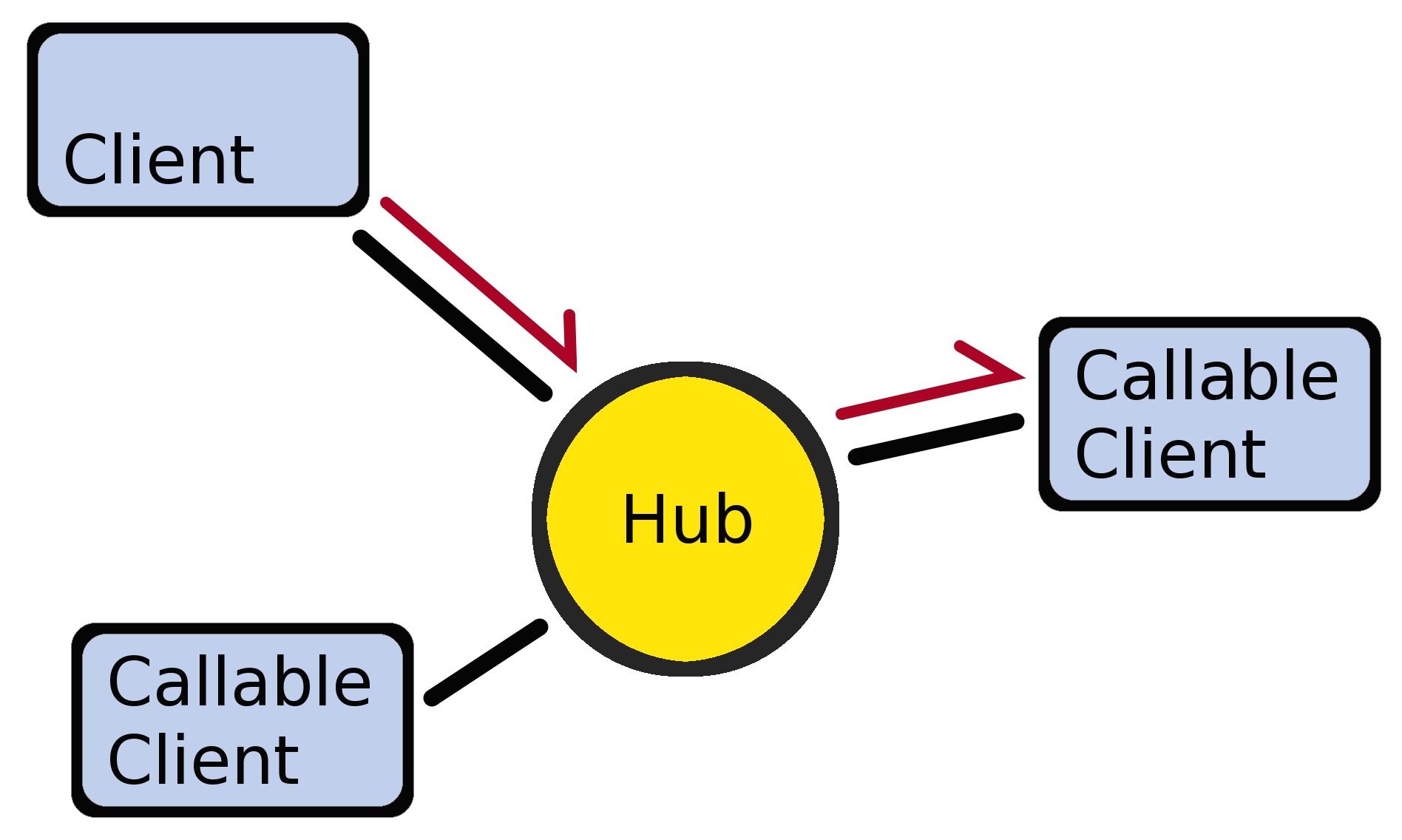}
\end{center}
\caption{\label{fig:topology}
Schematic of SAMP {\em Hub\/} and {\em clients\/} joined in a star topology.
Black lines indicate clients {\em registered\/} with the hub.
The red half-arrows indicate the progress of a message
from a sending client (which may or may not be {\em callable\/})
to a receiving client (which must be callable),
passing through the Hub.
}
\end{figure}

SAMP is based on a star topology, and its central component is a
{\em Hub\/} through which all communications are passed
(Figure \ref{fig:topology}).
Clients first perform a resource discovery step to locate the Hub,
and then {\em register\/} with it, establishing a private communication
channel through which subsequent calls to the Hub's services can be made.
These services include accepting metadata about the registering client,
providing information about other registered clients,
and forwarding {\em messages\/} to those clients.
These messages may elicit responses, which may optionally be passed
back to the message sender, again via the Hub.
All registered clients are able to send messages in this way.
Any client may optionally declare itself {\em callable\/},
in which case it is also able to receive messages sent by others.
Callability is optional since it is more difficult to
achieve in client code, requiring some server-like capacity
on top of the ability to invoke Hub services,
and simple actions like sending an image or table
can be achieved without this requirement.
In addition to declaring itself callable, a client wishing to
receive messages must explicitly {\em subscribe}\footnote{
   The term ``subscription'' derives from the ``publish-subscribe''
   messaging pattern.  It may however be more helpful to think of
   subscription as {\em declaring support for\/} a particular message type.
}
to one or more {\em MTypes} (message types).
Every message is labelled with an MType,
and the Hub will only deliver messages to clients
that have declared their interest in the MType in question with
an appropriate subscription.  When sending messages, clients may
either {\em broadcast\/} them to all subscribed clients or
target them to a named client, but in the latter case delivery
will fail if the target client has not appropriately subscribed.
If a client has no further use for SAMP communications
(for instance on application exit), it can and should {\em unregister\/}.

This framework combines the notions of publish-subscribe
(pub/sub) and Remote Procedure Call (RPC) messaging.
Like publish-subscribe, messages are only delivered to
appropriately subscribed recipients,
but like RPC the sender may optionally
target messages to a selected recipient,
and may optionally receive responses from the recipient(s).
The targetting mode, response requirement, and message content
are all decoupled from each other.

The details of this system are codified in a three-layer architecture:
\begin{description}
\item[Abstract API:] defines the services provided by the Hub and clients
\item[Profile:] maps the Abstract API to specific communication operations,
  such as bytes on the wire
\item[MTypes:] provide semantics for the actual messages exchanged between
  clients
\end{description}

Note that SAMP thus defines two distinct sets of
Remote Procedure Call (RPC) operations:
the functions declared by the Abstract API,
concerning the mechanics of client-hub communication and message delivery,
and SAMP Messages themselves classified by MType,
bearing the application-level content that clients wish
to exchange with each other.
The syntax and semantics of the former are carefully defined by the
SAMP standard, but the form and content of the latter are
agreed outside of SAMP itself by cooperating client developers.

Because of the central r\^{o}le of the Hub in this pattern,
it presents a single point of failure and potential bottleneck.
However, SAMP messages are usually short,
and in practice performance issues have not generally been apparent.

The following subsections present a more detailed account of
these ideas, along with some of the considerations that influenced
their design.
Sections \ref{sec:abstractApi}, \ref{sec:profile} and \ref{sec:mtypes}
describe the three architectural layers listed above,
and sections \ref{sec:typeSystem} and \ref{sec:extVocabs}
describe the underlying type system and how it is used to underpin
extensibility in SAMP.

\subsection{Abstract API} \label{sec:abstractApi}

The Abstract API defines the messaging capabilities of SAMP.
It takes the form of a list of a dozen or so function definitions
with typed arguments and return values, and well-defined semantics.
Most of these functions represent services provided by the Hub,
for instance
{\tt register} (which returns information required by the
client for future communications, typically an identification token)
and {\tt notifyAll} (which requests forwarding a given message to
all appropriately subscribed clients).
The remainder represent services required from callable clients, such as
{\tt receiveNotification} (which consumes a given message originating
from another client).

The messaging model in principle associates
a response with every message, containing at least a completion
status flag along with zero or more MType-defined return values.
However it is up to the sending client whether a response is
required from any given message; in many cases the status flag
is the only return value, and in this case a sending application
may or may not wish to make the effort to pass this on to its user
(for instance ``the table you sent was/was not successfully received'').
If the sender has no interest in the return value, it can use
the ``send-and-forget'' ({\em notification\/}) pattern,
with lower cost for sender, recipient and hub.

Message processing is fundamentally asynchronous from the
receiver's point of view, so that message/response times are
not limited to the lifetime of an RPC call in the underlying
transport mechanism.
However, the Hub provides an optional synchronous
fa\c{c}ade for sending messages when clients expect fast turnaround
and wish to avoid the additional complication of asynchronous processing.

\subsection{Profile} \label{sec:profile}

A particular SAMP Profile is what
turns the Abstract API into a set of rules
that a client can actually use to communicate with a running Hub,
and hence with other clients.

It performs two main jobs: first, it describes
how the functions defined by the API are turned into concrete
communication operations, by specifying an RPC-capable
transport mechanism and rules for mapping the SAMP
data types into the parameters and responses used by that mechanism.
Second, it defines a hub discovery mechanism,
which tells clients how to establish initial communications with the Hub,
usually involving some authentication step.
Particular profiles may also specify additional
profile-specific hub or client services exposed as functions
alongside those mandated by the Abstract API.

Initially (SAMP 1.11, 2009)
only a single profile was defined, the Standard Profile.
This uses XML-RPC\footnote{
   XML-RPC is a simple protocol for remote procedure calling
   based on HTTP and XML. It resembles a very much slimmed-down SOAP.
   Documentation can be found at \url{http://www.xmlrpc.com/}.
}
as a transport mechanism,
and allows hub discovery by storing the URL of the hub's XML-RPC server
along with a secret randomly generated key in a private ``lockfile''
in the user's home directory.

Version 1.3 of the standard (2012) introduced a second, the Web Profile,
for use by web-based clients.
This is required for applications running within web pages,
since the sandboxed environment imposed by the browser
makes the Standard Profile inaccessible.
It shares use of XML-RPC and some other characteristics with the
Standard Profile, but hub discovery has to be done differently,
and there are a number of complications to do with security,
described in \citet{adassxxi_paper} as well as the Standard.

This decoupling between the functionality of the service interface
and its incarnation in a specific transport mechanism
allows different transports to be introduced without changes to the core
protocol or existing clients, and has a number of benefits.
In a given SAMP session, a client may use the most appropriate
Profile for its SAMP communications and exchange
messages seamlessly with other clients using different profiles;
a desktop application can exchange messages with a web page
just as easily as with another desktop client.
This works because clients only ever
communicate directly with the Hub and not with each other,
while the Hub performs lossless translation between
profile-specific network operations
and the messaging model defined by the Abstract API.

Future requirements may result in additional Profile definitions,
and there is nothing in principle to prevent hub developers
from implementing new ones outside the frame of the SAMP standard.
However, from an interoperability point of view it is important
that all profiles are supported by all common Hub implementations,
so that a client can rely on the availability of a chosen profile
in a SAMP environment, and for this reason unnecessary proliferation
of profiles is discouraged.


\subsection{MTypes} \label{sec:mtypes}

An MType (message type) is the description for a message with particular
syntax and semantics.
It is analogous to a function definition in an API,
and consists of a labelling string (sometimes itself also known as the MType)
along with
a set of zero or more typed and named arguments,
a set of zero or more typed and named return values,
and some associated semantics indicating what the sender of such a
message is trying to convey.

By way of example,
a commonly used MType is {\tt image{\kdot}load{\kdot}fits},
defined like this:
\begin{quote}
  {\em Name:\/} \\
    \hspace*{1em} {\tt image{\kdot}load{\kdot}fits} \\
  {\em Semantics:\/} \\
    \hspace*{1em} Loads a 2-d FITS image \\
  {\em Arguments:\/} \\
    \hspace*{1em} {\tt url} (string): \\
    \hspace*{2em} URL of the image to load \\
    \hspace*{1em} {\tt image-id} (string, optional): \\
    \hspace*{2em} Identifier for use in subsequent messages \\
    \hspace*{1em} {\tt name} (string, optional): \\
    \hspace*{2em} Name for labelling loaded image in UI \\
  {\em Return Values:\/} \\
    \hspace*{1em} None.
\end{quote}

The name is a short hierarchical string composed of atoms separated
by the ``.'' character.
As well as identifying to a recipient the type of an incoming message,
it is used by clients to {\em subscribe\/} to messages, that is to
indicate to the Hub which messages they are prepared to receive.
For the purpose of subscription a limited wildcarding syntax is available,
so by using the MType patterns
{\tt image{\kdot}load{\kdot}fits}, {\tt image.*} or {\tt *}
a client may declare interest in only the above message,
or all image-related messages, or all messages, respectively.

In general, a callable client will only subscribe to those MTypes on which
it can meaningfully act, so for instance an image analysis tool
typically would subscribe to {\tt image{\kdot}load{\kdot}fits},
but not to {\tt spectrum{\kdot}load{\kdot}ssa-generic}.
A client that has an image FITS file to send can then
either query the Hub for those clients subscribed to the image load
message and offer its user the choice of which one to target,
or request the Hub to {\em broadcast} the image load message
to all (and only) the image-capable clients.

\subsection{Type System} \label{sec:typeSystem}

Supporting the function list defined by the Abstract API
and the parameters and return values specified by MTypes
is a type system defining the types of value permitted,
as well as rules for encoding various structured objects
using these types: message objects themselves, success and failure
message responses, application metadata, and MType subscription lists.

This system contains only three types:
{\em string}, {\em list\/} and {\em map}.
A string is a sequence of 7-bit ASCII printable characters,
a list is an ordered sequence of strings, lists or maps, and
a map is an unordered set of associations of string keys with values
of type string, list or map.
Structured objects are specified by the use of well-known keys in maps,
there is no special representation for null values,
and non-string scalar types must be serialized as strings.
(Obvious) conventions are suggested for serializing
integer, floating point and boolean values into string form,
but these suggestions are provided for the convenience
of MType definitions that wish to exchange such values without
reinventing the wheel, and are not a normative part of the protocol.

This restricted type system has been deliberately chosen to introduce
minimal dependency of messaging behaviour on the details of
non-core parts of the delivery system,
in particular profile-specific transport mechanisms and language-specific
client libraries.
This both reduces the restrictions on what languages and transport layers
may be used with SAMP, and ensures that values transmitted will
not be modified during processing by parts of the messaging system
outside of client control.


The type system is rich enough to represent complex structured data
where required, but note it is not intended for use with
binary data, and transmission of bulk data or large payloads
in general is discouraged within SAMP messages in favour of
passing URLs around instead, meaning that
client and Hub implementations can work on the assumption
of short message payloads.

This convention of out-of-band bulk data transfer does place
an additional burden on sending clients however,
since to transmit a bulk data item (such as a table or image)
not already available from an existing URL
it is necessary to make it so available, for instance by writing
bytes to a temporary file or serving them from an embedded HTTP server.
It can also present complications if the sending and receiving client
are not able to see the same URLs, for instance due to different security
contexts; in this case, additional Hub services may be required to
assist with data transfer between domains
(accordingly, the Web Profile provides services to assist with
cross-domain data exchange).

Note also that the string type does not natively accommodate Unicode text,
including XML.
The restriction to 7-bit ASCII is driven by the requirement for use
from non-Unicode-capable environments such as Fortran, IDL and
some shell scripting languages.
This has not caused known problems to date, but inability to handle
Unicode text without additional encoding could prove awkward in
some cases, and it may be necessary to revisit this restriction in a future
revision of the standard.

\subsection{Extensible Vocabularies} \label{sec:extVocabs}

Extensibility is built into this system via the notion of an
{\em extensible vocabulary} used when representing structured objects.

Structured objects are represented as maps with well-known keys,
but the rule is that additional keys are always permitted,
and that hubs and clients must ignore any keys they do not understand,
propagating them to downstream consumers where applicable
(compare the NDF extension architecture described in \citet{sgp38}).
A corollary is that such non-well-known keys must be defined in such
a way that ignoring them will result in reasonable behaviour.
The Abstract API tends to prefer {\em maps\/} (unordered name/value pairs)
over ordered parameter value lists, which makes this extensibility
pervasive throughout the messaging system,
applying for instance to
client metadata and subscription declarations,
message transmission information,
and MType-specified message parameter lists and return values.

For instance, a client sending a message must pass it to the Hub
as a map with two required keys:
{\tt samp{\kdot}mtype} giving the MType label and
{\tt samp{\kdot}params} giving the MType-specified parameter list
(itself a map).
But a client may optionally insert additional non-standard key/value pairs
into that map, for instance using a non-standard key {\tt priority}
to associate a particular priority level with the message.
If the Hub happens to support this non-standard feature,
it is able to treat the message specially in view of this declaration;
in any case it will propagate the message to recipient clients
with the additional entry present, so if one of those supports
this feature then it may use the value in processing.
The same rule applies for instance to the MType-determined
message parameter list;
an MType like {\tt image{\kdot}load{\kdot}fits}
has a required parameter {\tt url},
but a sending client may add a non-standard parameter like
{\tt colormap} (or {\tt ds9{\kdot}colormap}) alongside the well-known ones
for the benefit of any client that happens to support it.
Clients can therefore piggy-back experimental or application-specific
instructions on top of generic messages to achieve more detailed control
where available, falling back to the baseline functionality if it is not.
Using this extensibility pattern, new or enhanced features of 
particular MTypes or of the protocol itself can be prototyped 
very easily, requiring no changes to the SAMP standard or infrastructure
implementation beyond those components actually using
the non-standard features,
and imposing no negative impact on existing messaging operations.
If they are found to be useful, they may be adopted in the future
as (most likely optional) well-known keys alongside the original ones.

Some associated namespacing rules apply.
Well-known keys defined by the SAMP standard are in the reserved {\tt samp}
namespace, meaning they begin with the string ``{\tt samp.}''.
When introducing non-standard keys it is not permitted to use this
namespace, but any other syntactically legal string is allowed.
The special namespace {\tt x-samp} is available
for keys proposed for future incorporation into the standard,
and hubs and clients should treat keys which differ 
only in the substitution of {\tt samp} for {\tt x-samp} as identical,
to ease standardisation of prototype features.
In the case of MType parameters and return values,
which are mostly not defined by SAMP itself, there is no reserved namespace.



\section{Use in Practice} \label{sec:usage}

The protocol described above is capable of supporting a wide range of
different messaging patterns.
For use in a particular scenario,
a number of practical considerations must also be worked out.
This section discusses how the framework has been applied to date
to support the original goal of helping to integrate data analysis
tools used by astronomers.

Section \ref{sec:hubProvision} explains how hub provision is managed,
section \ref{sec:messageSemantics} describes some common patterns of
message semantics, and
section \ref{sec:mtypeProcess} addresses the social mechanisms by which
these are agreed on by the SAMP community.
Section \ref{sec:software} reviews the existing landscape of
SAMP infrastructure software and SAMP-aware tools, and
Section \ref{sec:examples} provides some concrete examples of it in action.

\subsection{Hub Provision} \label{sec:hubProvision}

SAMP's star topology means that a Hub (in most cases, exactly one Hub)
must be running for any messaging to take place.
Ideally, an independent Hub process would be started
as part of user session setup to ensure its constant availability.
While this is quite possible and appropriate in some scenarios,
even the minimal configuration required to establish it
(a hub startup line in a session startup file)
requires the kind of explicit user action which cannot always be relied upon.
Simply put, if the functionality doesn't show up in the user interface
with zero user effort, most users will never discover it.

It is common practice therefore, though by no means a requirement,
for some SAMP-aware tools to come with an embedded hub.
In this scenario, when a SAMP-aware tool starts up, it first checks for
a running hub.
If one exists, it registers with it;
if not, and if it has the capability, it starts its own embedded Hub,
and registers with that.
Note that a client running an embedded Hub communicates with it in
just the same way as with an independent one, it has no privileged access.
Non-hub-capable clients may choose to check for a running Hub and connect
on startup, on explicit user request, or when periodic polling indicates
that one has become available.
The effect is that usually when two or more SAMP-aware tools are running,
a Hub will be present and those tools will find themselves connected to it,
enabling messaging.
Sometimes the application hosting the embedded hub will be shut down
during a session taking the Hub down with it,
and in that case another application may notice the fact and start one up,
at which point some or all previously registered clients may notice
the new hub and re-register with it.

This somewhat haphazard model of hub provision does not form a
robust platform for high-reliability messaging,
but, in accordance with SAMP's pragmatic approach,
operates well enough most of the time, with a minimum of user effort;
usually, it ``just works''.
Note that where explicit control of an independent hub process
can be arranged,
for example as part of a managed user environment,
more robust connectivity will result;
an example is the Herschel Interactive Processing Environment
\citep{2012ASPC..461..733B}.

\subsection{MType Semantics} \label{sec:messageSemantics}

A messaging framework only serves any purpose if there exists
a vocabulary of messages understood by the
applications which are going to exchange them.
In SAMP terms that means establishing a collection of
more or less well-known MTypes (section \ref{sec:mtypes}).
Choosing the right semantics for this collection is crucial
to the utility and character of the messaging system in practice.

The most obvious approach for providing message-based control
of an application
is to identify (at least some of) the capabilities it offers and
define a messaging interface with parameters and return values
exposing those capabilities.  An image display application might
expose a set of MTypes allowing image load into a new window,
zoom configuration, colour map choice, WCS display and so on.
This allows other applications to control its behaviour
in detail and is suitable for tight integration of a known set
of tools with a good understanding of each other's capabilities,
for instance to execute a pre-orchestrated sequence of processing steps.

However, this approach is less effective in less predictable environments.
The controlling client needs to understand
the capabilities of its partner client in order to control it.
But if the set of tools in use at runtime is chosen by a
user from an open-ended set
rather than mandated by a developer, the identity of the
partner client or clients is not known in advance.
In general, different applications even of similar types
have different capability sets
and internal data models, and these cannot readily be encompassed
by any single general abstraction.
Different image display tools
may support different data formats,
may or may not support multiple loaded windows or images,
may specify zooms in different ways,
may offer different selections of colour maps,
may provide WCS display with different options or not at all
and so on,
and the burden on a client wishing to control a range
of different recipients quickly becomes large.
Even if an application developer is prepared to study the
messaging APIs offered by existing available tools and implement logic
managing message dispatch for each case,
the resulting code will not cope with applications
unknown to the developer,
for instance ones yet to be written.

For uncontrolled environments in which the user selects
the range of cooperating tools at runtime therefore,
a ``loose integration'' model has turned out to be more successful.
This approach focuses on a messaging interface
consisting of a fairly small number of MTypes with
semantics that are non-client-specific and rather vague.
The semantics of the most-used messages generally boils down to
``{\em Here is an X}'', where $X$ may be some resource type such as a
table, image, spectrum,
sky position, coverage region, bibcode etc,
or sometimes a reference to an $X$ from a previous message,
for instance a row selection relating to an earlier-sent table.
The implication of such an MType is that the receiver should do
something appropriate with the $X$ in question: load, display, highlight,
or otherwise perform some action which makes sense given the
receiving application's capabilities.
Callable SAMP clients should therefore advertise themselves
(by subscribing to the appropriate MTypes) as $X$-capable
tools only if they are in a position to do something sensible
with an $X$ should they be presented with one.
Such an advertisement serves as a hint to potential $X$-senders,
though it does not constitute a guarantee of any particular behaviour.
This framework typically manifests itself in a client user interface
as an option, for an $X$ currently known by that client,
either to {\em broadcast\/} it to all $X$-capable clients,
or to target it to an entry selected by the user
from a dynamically-discovered list of $X$-capable clients
(see Figure \ref{fig:linked} for an example).
For this kind of usage, the presence of a human in the loop
to direct messages between clients as required by a particular workflow
is an important part of the system,
but the decisions required from the user are generally simple ones,
e.g.\ which of a small list of clients to contact.

For clients to interoperate as reliably as possible in this scenario,
it is not sufficient just to agree on the notion of a table or an image
for exchange, it is also necessary to specify the exact data exchange
format.
In the case of tabular data, a variety of possible exchange formats
is in common use: FITS binary and ASCII tables, VOTable,
Comma-Separated Values and a host of others including 
many ASCII-based variants.  Different choices are convenient in
different usage contexts, suggesting the need for a variety of
distinct format-specific MTypes.
However, a proliferation of alternative exchange formats,
though superficially convenient, erodes interoperability.
If multiple exchange formats capaple of serializing the same thing
are available,
the sender has to choose which to send, and the receiver may or
may not be able to receive it.
Well-behaved recipients should include conversion code for as many
formats as possible, and well-behaved senders should send data
in a format dependent on what is supported by the intended recipient.
For applications willing to expend a lot of effort on interoperability 
the work required at both ends increases rapidly with the number of
available formats,
while less conscientious implementations may find themselves unable to
exchange data of essentially compatible types,
or the community of SAMP clients may fragment into format-specific
sub-communities unable to communicate globally.
As much as possible therefore, it is desirable to restrict the options
to a single well-defined exchange format for each basic data type.

This can be a difficult balance to get right.
In the case of images, astronomy is fortunate that FITS serves as
the {\em lingua franca\/}, and the only commonly-used MType is
{\tt image{\kdot}load{\kdot}fits}.
For tabular data, clients are strongly encouraged to use
{\tt table{\kdot}load{\kdot}votable} even if it means translating to/from
some other format;
however other {\tt table{\kdot}load.*} variants are in use
for specialist purposes,
for instance for the CDF format\footnote{\url{http://cdf.gsfc.nasa.gov/}},
which though tabular is
not readily translated to VOTable without loss of information,
and which tends to be used in communities not familiar with VOTable.
In the case of spectra, for various reasons related to the
form in which spectral data is typically obtained and
the typical capabilities of spectrum-capable clients,
the relevant MType is {\tt spectrum{\kdot}load{\kdot}ssa-generic},
which permits any format to be used for the spectral data,
with additional parameters specifying
the format actually in use.

SAMP is capable of supporting both tight and loose integration,
and both are in use,
but for coupling interactive data analysis tools the loose integration
model has proven the most productive,
and able to support ways of working that have not been possible
using other available messaging systems.

\subsection{MType Definition Process} \label{sec:mtypeProcess}

A suitable collection of MTypes must not only exist but be known by
potential message senders -- which means the authors of the relevant
software -- in order for useful messaging to take place.

In the case of application-specific MTypes, the documentation of
available MTypes and their definitions is clearly
best handled as part of the documentation of the application itself.
These typically provide functionality that only makes sense for
a given tool, and make use of a suitably specific namespace,
for instance {\tt script{\kdot}aladin{\kdot}send},
which allows external applications to control Aladin
by sending commands in its scripting language.

Developers are also free to define their own MTypes for use privately
or in some closed group with locally agreed conventions for documentation,
perhaps to support some tight-coupling-like usage.

However, for well-known MTypes intended for unrestricted use,
for instance of the loose-coupling variety described above,
some public process is required to establish
and publicise their definitions, so that client developers can
both become aware of the conventions currently in use by other tools,
and contribute their requirements for new or modified functionality.

One possibility
is to decide on a fixed list to form part of the SAMP standard.
A small number of ``administrative'' MTypes,
concerned with the messaging infrastructure,
for instance {\tt samp{\kdot}hub{\kdot}event{\kdot}register}
which informs existing clients when a new client has registered,
have been written in to the standard in this way.
All of these are in the reserved {\tt samp.} namespace.
However, for astronomy-specific MTypes this option was rejected,
partly in order to avoid the introduction of astronomy-specific details
into a standard which is otherwise not tied to a particular domain,
and partly because the rather heavyweight IVOA process for
standard review \citep{ivoadoc},
in which draft to acceptance rarely takes less than 12 months,
would impede introduction and updating of MTypes as required
by implementation experience and new application demands.
Another option is periodic publication of MType definitions
in an IVOA Note.  Such Notes may be issued at will without formal review,
but no straightforward updating mechanism is in use,
and this option was still felt to be undesirably cumbersome.

Instead, a wiki page\footnote{
  At time of writing, this wiki page can be found near
  \mbox{\url{http://www.ivoa.net/samp/}}.
  If the process for MType publication changes in the future,
  that URL should still indicate where to look for a list.}
was set up on the IVOA web site listing currently agreed MTypes.
An informal understanding was adopted in which
application developers are encouraged to discuss requirements for
new MTypes or modifications to existing ones either privately or
on the associated mailing list\footnote{apps-samp@ivoa.net}, 
and if consensus is reached, to edit the wiki page accordingly.
This was intended as a provisional measure to be reviewed and
modified as required, but, six years later, the need for a more
formal process has not been apparent, and there are no current
plans to modify this arrangement.

At time of writing, a dozen MTypes are listed on the wiki,
concerned with exchanging tables, row selections, FITS images, spectra,
sky coordinates, VO Resource identifiers \citep{ACVOregistry},
MOC sky coverage maps \citep{moc_std},
bibcodes and one or two other items.
The list has been fairly stable, though new entries and new
optional parameters are sometimes added as required.

\subsection{Existing Software} \label{sec:software}

Since its first version in 2008, a wide range of SAMP-enabled
infrastructure and application software has become available.

Of infrastructure sofware that is actively maintained at the time of writing,
interchangeable Hub implementations exist in Java and Python,
and client toolkits in Java, Python, C and Javascript.
Validation, debugging and development support tools are also available.
Historical, partial or experimental SAMP functionality has
also appeared in other languages including Perl, C Sharp and IDL.

Applications using SAMP number in the dozens, and include
GUI analysis tools for images, catalogues, spectra, SEDs,
time series and interferometry data,
observation tools,
outreach applications,
command-line and graphical data access and manipulation suites,
interactive processing environments,
web archives either exposing simple results pages
or offering sophisticated browser functionality,
throwaway user scripts,
and more.

SAMP infrastructure libraries are surveyed in \citet{adassxxii_bof},
though a notable more recent development is the inclusion of the Python
hub and client implementation in the Astropy package
\citep{2013A&A...558A..33A} since its version 0.4.
A partial list of some other tools with SAMP functionality
may also be found near \url{http://www.ivoa.net/samp}.

While initially developed and used mostly
within stellar and galactic astronomy,
use is now becoming common in related fields such as
planetary science \citep{ACVOplanetary}
and space physics \citep{ACVOspacephys}.
It is probably now the case that most astronomical applications
that can benefit from interaction with other such tools
include at least a basic SAMP capability.
It is harder to ascertain to what extent this functionality is used
in practice, but the enthusiasm of application developers to
incorporate SAMP is presumably indicative of its utility.

\subsection{Examples of Use} \label{sec:examples}

\begin{figure*}
\begin{center}
\begin{tabular}{lc}
{\em (a)} & \\
& \includegraphics[width=0.92\textwidth]{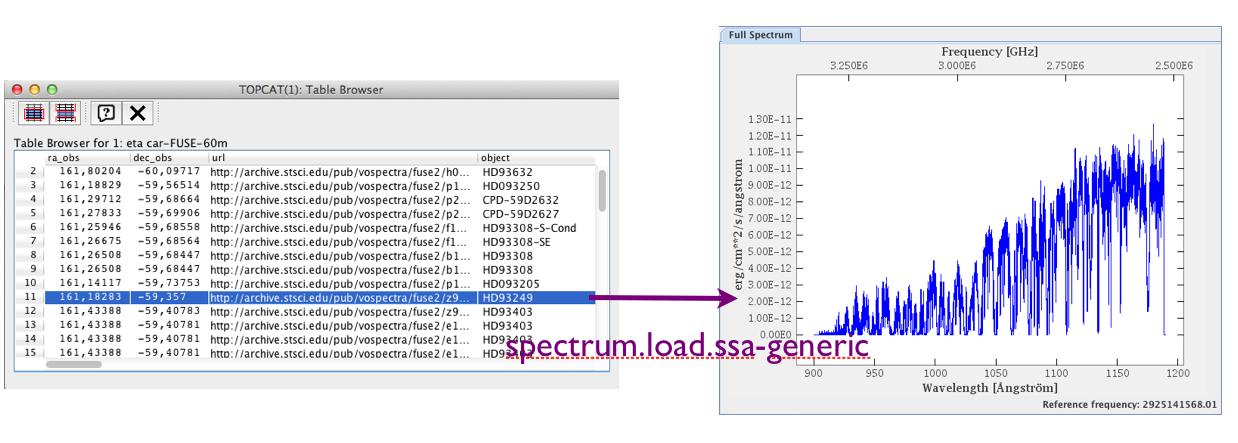} \\
{\em (b)} & \\
& \includegraphics[width=0.92\textwidth]{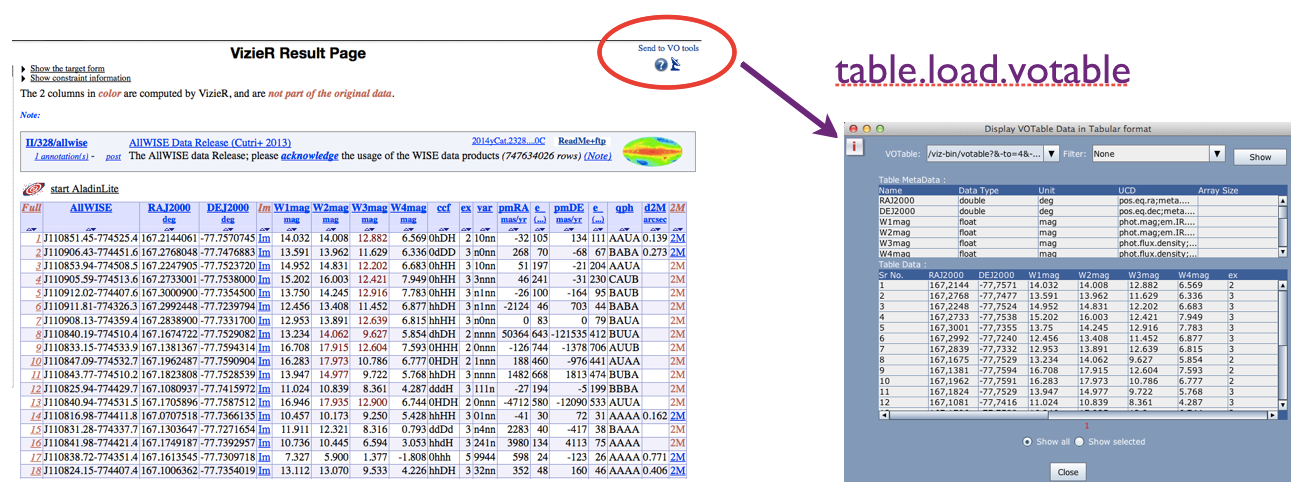} \\[2mm]
{\em (c)} & \\
& \includegraphics[width=0.92\textwidth]{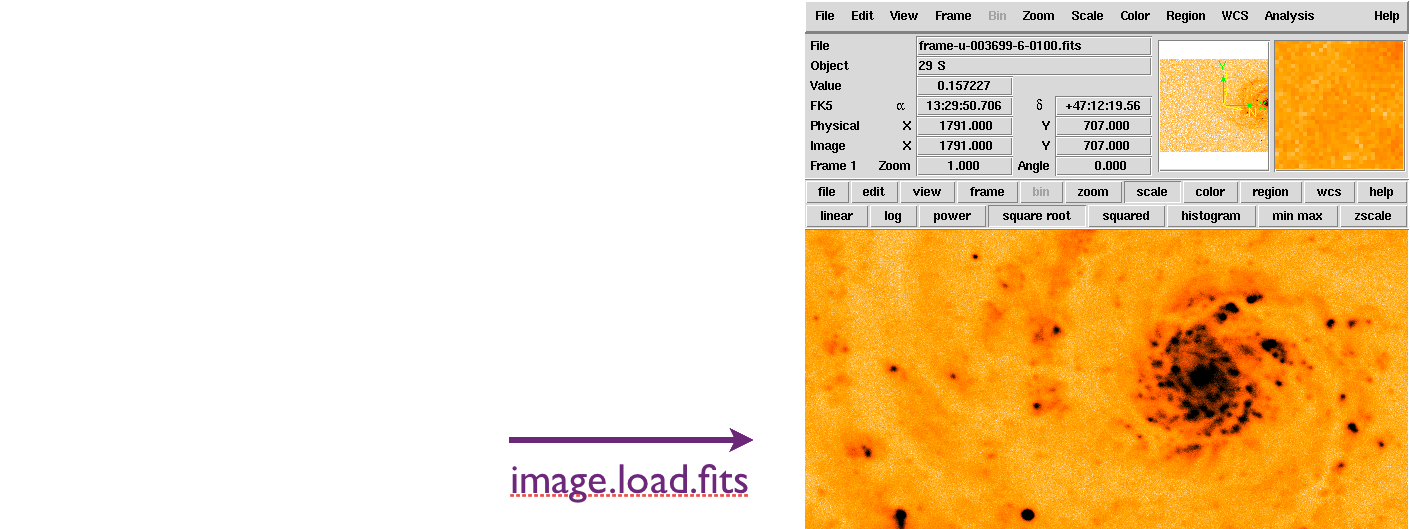} \\
\begin{picture}(0,0)
\put(0,0){
\hspace*{5mm}
{\small\tt\color{olive}
\raisebox{18mm}{
\begin{minipage}[b]{0cm}
\begin{tabbing}
{\color{gray}>>>} {\color{violet}from} {\color{darkgray}astropy.vo.samp} {\color{violet}import} {\color{darkgray}SAMPIntegratedClient} \\
{\color{gray}>>>} {\color{darkgray}client} = {\color{darkgray}SAMPIntegratedClient}() \\
{\color{gray}>>>} {\color{darkgray}client}.{\color{darkgray}connect}() \\
{\color{gray}>>>} {\color{darkgray}client}.{\color{darkgray}notify\_all}(\{ \\
{\color{gray}...} \ \ \ {\color{blue}'samp.mtype'}:\ {\color{blue}'image.load.fits'}, \\
{\color{gray}...} \ \ \ {\color{blue}'samp.params'}:\ \{ \\
{\color{gray}...} \ \ \ \ \ \ {\color{blue}'url'}:\ {\color{blue}'file:///data/ngc5194/'} \\
{\color{gray}...} \ \ \ \ \ \ \ \ \ \ \ \ \ {\color{blue}'frame-u-003699-6-0100.fits'}, \\
{\color{gray}...} \ \ \ \ \ \ {\color{blue}'name'}:\ {\color{blue}'SDSS band u'} \\
{\color{gray}...} \ \ \ \} \\
{\color{gray}...} \})
\end{tabbing}
\end{minipage}}}
\hspace*{2mm}
}
\end{picture}
\end{tabular}
\end{center}
\caption{\label{fig:transmit}
Transmission of data items from one SAMP client to another.
(a) A table has been acquired using Simple Spectral Access Protocol
by the TOPCAT table analysis tool, in which each row references
an external spectrum by URL.  When the user selects a row of this
table, the spectrum is sent to the CASSIS spectrum tool.
(b) The source catalogue resulting from a VizieR query,
displayed in a web page, is sent to the VOPlot application.
(c) A local FITS image is loaded into the ds9 image viewer
from the AstroPy command line.
}
\end{figure*}
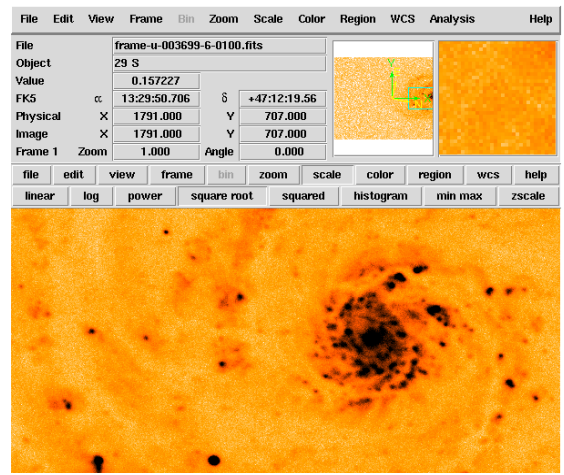

\begin{figure*}
\begin{center}
\includegraphics[height=6cm]{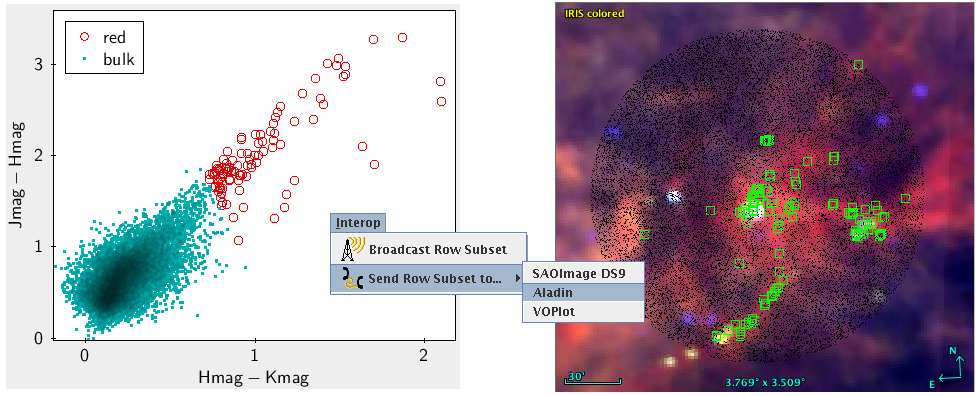}
\end{center}
\caption{\label{fig:linked}
Linked view of data in TOPCAT and Aladin showing
red objects in the vicinity of V$^{\star}$~V410~Tau.
A full list of sources in the region has been loaded
into Aladin (right), then transferred to TOPCAT (left),
using SAMP MType {\tt table.load.votable}.
The user plots a colour-colour diagram in TOPCAT
and selects the reddest objects graphically,
causing them to be displayed as red circles,
then passes the selection back to Aladin
(MType {\tt table.select.rowList}) where
they are shown as green squares;
the inset menu shows TOPCAT's user interface for this step.
Clicking on one of the points in either application can
then highlight the corresponding point in the other
(MType {\tt table.highlight.row}).
This example illustrates how SAMP enables seamless exploration of
data using a combination of parameter space and physical space.
}
\end{figure*}

Current SAMP usage is most prevalent along the lines of the scenario
outlined in the Introduction, allowing desktop and web-based clients to
cooperate as a loosely integrated suite
by employing the small number of data-exchange MTypes in common use.
Figure \ref{fig:transmit} shows some examples of
the basic case where one client transmits a data item
(spectrum, table or image) to another.
Figure \ref{fig:linked} illustrates a more sophisticated
interaction in which two applications exchange data and control
in both directions to provide linked views exploiting the
capabilities of each.

SAMP has also been employed in other ways however,
for instance to provide a private layer for RPC functionality
required internally by Iris \citep{ACVOiris}
and to experiment with visualisation using on-demand
data generation in Astro-WISE \citep{2013ExA....35..283B}.

\section{Conclusions} \label{sec:conclusion}

SAMP provides a flexible and easy to use messaging framework,
deployed in much current astronomical software,
which supports various models of inter-tool communication.

The most productive of these models to date has been loose coupling
between a user-selected set of independently developed
interactive data acquisition and analysis applications,
to deliver functionality approaching that of an
integrated suite.
This model is built on SAMP's combination of
publish-subscribe messaging,
vague message semantics,
and ease of adoption by both developers and users
leading to widespread uptake.

SAMP's flexibility means that it is capable of supporting other
communication models, some in more marginal use now
and some which may be explored further in the future.
Introducing new Profiles, different MType libraries or
alternative hub provision arrangements could render the same
infrastructure suitable for contexts with different
requirements for reliability, security or scalability.
Another possible scenario is inter-host messaging to
support collaborative work;
this option has been under consideration throughout SAMP's history
and is possible using existing profiles,
though in current configurations
it is somewhat cumbersome to set up and has so far not
received much user attention.
Despite its development history, there is nothing in the protocol
specific to either the Virtual Observatory or astronomy,
so use in other problem domains is quite feasible,
though the authors are not aware of effort currently being
deployed in this direction.

SAMP's design has been informed by the requirements and experimentation
of the SAMP developer community, largely within the context of
the Virtual Observatory,
including positive and negative lessons learned from its predecessor PLASTIC.
Some aspects of this design that have proved particularly successful
include
the decoupling of architectural concerns into API, transport
mechanism and semantics,
the lightweight, bottom-up process for agreement of semantics,
and the built-in extensibility provided by pervasive use of
extensible vocabularies.

Together, these fall under the heading of standardising
only those things which need to be defined at a given stage,
and leaving the option of filling in the details until a time
and in a context when the requirements will be clearer.
The need for the Web Profile was not forseen when the first version
of the standard was published, but the transport/API decoupling
meant it could be retrofitted with no disruption to existing client code.
The fact that MType semantics are excluded from the standard
itself means that these can be defined and iteratively adjusted with
experience of a working transport infrastructure, rather than
specifying them up front by committee decision as part of the
protocol design, only to find them ill-adapted to tool
deployment in practice.

Other factors important to its success have been
the small number of MTypes actually in common use,
enabled by the convention of vague message semantics
and standardisation on data formats,
and the unobtrusive embedding of SAMP into existing applications
meaning that the functionality is available without
requiring any special setup or understanding from the user.
The IVOA and other cross-institutional forums associated with
the Virtual Observatory movement have also been of considerable
importance in enabling and encouraging the necessary communication
between application developers, though much software from outside
the VO community is now also involved.

SAMP is not a magic bullet.
In typical current use the level of integration it offers
between independently developed tools falls short of what would be
available from a monolithic application,
its pragmatic approach to communications can lead to patchy reliability,
and its security model would not be appropriate for use with
commercially sensitive data.
However its ease of use and widespread uptake have delivered in practice
an improved environment for desktop data analysis,
allowing working astronomers to get more done.

\section*{Acknowledgements}

The development and adoption of SAMP has been a collaborative effort
between participating developers within the Virtual Observatory
community, particularly the Applications Working Group of the IVOA.
Initial work on PLASTIC and SAMP was supported by the EU FP6
project VOTech and FP7 project AIDA.  The SC4DEVO workshop series
was funded by the UK e-Science Core Programme.
MBT has additionally been supported in development
of the SAMP protocol and software
by several grants from the UK's PPARC and STFC research councils,
and in writing this paper by STFC grant ST/J001414/1.
Development of the SAMP Web Profile was supported by Microsoft Research.
All of this support is gratefully acknowledged.

\section*{References}

\bibliography{bibsamp}

\begin{thebibliography}{24}
\expandafter\ifx\csname natexlab\endcsname\relax\def\natexlab#1{#1}\fi
\providecommand{\url}[1]{\texttt{#1}}
\providecommand{\href}[2]{#2}
\providecommand{\path}[1]{#1}
\providecommand{\DOIprefix}{doi:}
\providecommand{\ArXivprefix}{arXiv:}
\providecommand{\URLprefix}{URL: }
\providecommand{\Pubmedprefix}{pmid:}
\providecommand{\doi}[1]{\href{http://dx.doi.org/#1}{\path{#1}}}
\providecommand{\Pubmed}[1]{\href{pmid:#1}{\path{#1}}}
\providecommand{\bibinfo}[2]{#2}
\ifx\xfnm\relax \def\xfnm[#1]{\unskip,\space#1}\fi
\bibitem[{{Astropy Collaboration} et~al.(2013){Astropy Collaboration},
  {Robitaille}, {Tollerud}, {Greenfield}, {Droettboom}, {Bray}, {Aldcroft},
  {Davis}, {Ginsburg}, {Price-Whelan}, {Kerzendorf}, {Conley}, {Crighton},
  {Barbary}, {Muna}, {Ferguson}, {Grollier}, {Parikh}, {Nair}, {Unther},
  {Deil}, {Woillez}, {Conseil}, {Kramer}, {Turner}, {Singer}, {Fox}, {Weaver},
  {Zabalza}, {Edwards}, {Azalee Bostroem}, {Burke}, {Casey}, {Crawford},
  {Dencheva}, {Ely}, {Jenness}, {Labrie}, {Lim}, {Pierfederici}, {Pontzen},
  {Ptak}, {Refsdal}, {Servillat} and {Streicher}}]{2013A&A...558A..33A}
\bibinfo{author}{{Astropy Collaboration}}, \bibinfo{author}{{Robitaille},
  T.P.}, \bibinfo{author}{{Tollerud}, E.J.}, \bibinfo{author}{{Greenfield},
  P.}, \bibinfo{author}{{Droettboom}, M.}, \bibinfo{author}{{Bray}, E.},
  \bibinfo{author}{{Aldcroft}, T.}, \bibinfo{author}{{Davis}, M.},
  \bibinfo{author}{{Ginsburg}, A.}, \bibinfo{author}{{Price-Whelan}, A.M.},
  \bibinfo{author}{{Kerzendorf}, W.E.}, \bibinfo{author}{{Conley}, A.},
  \bibinfo{author}{{Crighton}, N.}, \bibinfo{author}{{Barbary}, K.},
  \bibinfo{author}{{Muna}, D.}, \bibinfo{author}{{Ferguson}, H.},
  \bibinfo{author}{{Grollier}, F.}, \bibinfo{author}{{Parikh}, M.M.},
  \bibinfo{author}{{Nair}, P.H.}, \bibinfo{author}{{Unther}, H.M.},
  \bibinfo{author}{{Deil}, C.}, \bibinfo{author}{{Woillez}, J.},
  \bibinfo{author}{{Conseil}, S.}, \bibinfo{author}{{Kramer}, R.},
  \bibinfo{author}{{Turner}, J.E.H.}, \bibinfo{author}{{Singer}, L.},
  \bibinfo{author}{{Fox}, R.}, \bibinfo{author}{{Weaver}, B.A.},
  \bibinfo{author}{{Zabalza}, V.}, \bibinfo{author}{{Edwards}, Z.I.},
  \bibinfo{author}{{Azalee Bostroem}, K.}, \bibinfo{author}{{Burke}, D.J.},
  \bibinfo{author}{{Casey}, A.R.}, \bibinfo{author}{{Crawford}, S.M.},
  \bibinfo{author}{{Dencheva}, N.}, \bibinfo{author}{{Ely}, J.},
  \bibinfo{author}{{Jenness}, T.}, \bibinfo{author}{{Labrie}, K.},
  \bibinfo{author}{{Lim}, P.L.}, \bibinfo{author}{{Pierfederici}, F.},
  \bibinfo{author}{{Pontzen}, A.}, \bibinfo{author}{{Ptak}, A.},
  \bibinfo{author}{{Refsdal}, B.}, \bibinfo{author}{{Servillat}, M.},
  \bibinfo{author}{{Streicher}, O.}, \bibinfo{year}{2013}.
\newblock \bibinfo{title}{{Astropy: A community Python package for astronomy}}.
\newblock \bibinfo{journal}{Astron Astrophys} \bibinfo{volume}{558},
  \bibinfo{pages}{A33}.
\newblock \DOIprefix\doi{10.1051/0004-6361/201322068},
  \href{http://arxiv.org/abs/1307.6212}{\tt arXiv:1307.6212}.
\bibitem[{{Balm}(2012)}]{2012ASPC..461..733B}
\bibinfo{author}{{Balm}, P.}, \bibinfo{year}{2012}.
\newblock \bibinfo{title}{{Herschel Interactive Processing Environment (HIPE):
  Open to the World and the Future}}, in: \bibinfo{editor}{{Ballester}, P.},
  \bibinfo{editor}{{Egret}, D.}, \bibinfo{editor}{{Lorente}, N.P.F.} (Eds.),
  \bibinfo{booktitle}{Astronomical Data Analysis Software and Systems XXI}, p.
  \bibinfo{pages}{733}.
\bibitem[{Boch et~al.(2006)Boch, Comparato, Taylor, Taylor and
  Winstanley}]{plastic_note}
\bibinfo{author}{Boch, T.}, \bibinfo{author}{Comparato, M.},
  \bibinfo{author}{Taylor, J.}, \bibinfo{author}{Taylor, M.},
  \bibinfo{author}{Winstanley, N.}, \bibinfo{year}{2006}.
\newblock \bibinfo{title}{PLASTIC - A Protocol for Desktop Application
  Interoperability}.
\newblock \bibinfo{type}{Technical Report}. IVOA Note.
\bibitem[{Boch et~al.(2014)Boch, Donaldson, Durand, Fernique, O'Mullane,
  Reinecke and Taylor}]{moc_std}
\bibinfo{author}{Boch, T.}, \bibinfo{author}{Donaldson, T.},
  \bibinfo{author}{Durand, D.}, \bibinfo{author}{Fernique, P.},
  \bibinfo{author}{O'Mullane, W.}, \bibinfo{author}{Reinecke, M.},
  \bibinfo{author}{Taylor, M.B.}, \bibinfo{year}{2014}.
\newblock \bibinfo{title}{{MOC -- HEALPix Multi-Order Coverage Map}}.
\newblock \bibinfo{type}{Technical Report}. IVOA Recommendation.
\bibitem[{{Bonnarel} et~al.(2000){Bonnarel}, {Fernique}, {Bienaym{\'e}},
  {Egret}, {Genova}, {Louys}, {Ochsenbein}, {Wenger} and
  {Bartlett}}]{2000A&AS..143...33B}
\bibinfo{author}{{Bonnarel}, F.}, \bibinfo{author}{{Fernique}, P.},
  \bibinfo{author}{{Bienaym{\'e}}, O.}, \bibinfo{author}{{Egret}, D.},
  \bibinfo{author}{{Genova}, F.}, \bibinfo{author}{{Louys}, M.},
  \bibinfo{author}{{Ochsenbein}, F.}, \bibinfo{author}{{Wenger}, M.},
  \bibinfo{author}{{Bartlett}, J.G.}, \bibinfo{year}{2000}.
\newblock \bibinfo{title}{{The ALADIN interactive sky atlas. A reference tool
  for identification of astronomical sources}}.
\newblock \bibinfo{journal}{Astron. Astrophys. Suppl. Ser.}
  \bibinfo{volume}{143}, \bibinfo{pages}{33--40}.
\newblock \DOIprefix\doi{10.1051/aas:2000331}.
\bibitem[{{Buddelmeijer} and {Valentijn}(2013)}]{2013ExA....35..283B}
\bibinfo{author}{{Buddelmeijer}, H.}, \bibinfo{author}{{Valentijn}, E.A.},
  \bibinfo{year}{2013}.
\newblock \bibinfo{title}{{Query driven visualization of astronomical
  catalogs}}.
\newblock \bibinfo{journal}{Experimental Astronomy} \bibinfo{volume}{35},
  \bibinfo{pages}{283--300}.
\newblock \DOIprefix\doi{10.1007/s10686-011-9263-0},
  \href{http://arxiv.org/abs/1110.2294}{\tt arXiv:1110.2294}.
\bibitem[{{Comparato} et~al.(2007){Comparato}, {Becciani}, {Costa}, {Larsson},
  {Garilli}, {Gheller} and {Taylor}}]{2007PASP..119..898C}
\bibinfo{author}{{Comparato}, M.}, \bibinfo{author}{{Becciani}, U.},
  \bibinfo{author}{{Costa}, A.}, \bibinfo{author}{{Larsson}, B.},
  \bibinfo{author}{{Garilli}, B.}, \bibinfo{author}{{Gheller}, C.},
  \bibinfo{author}{{Taylor}, J.}, \bibinfo{year}{2007}.
\newblock \bibinfo{title}{{Visualization, Exploration, and Data Analysis of
  Complex Astrophysical Data}}.
\newblock \bibinfo{journal}{The Publications of the Astronomical Society of the
  Pacific} \bibinfo{volume}{119}, \bibinfo{pages}{898--913}.
\newblock \DOIprefix\doi{10.1086/521375},
  \href{http://arxiv.org/abs/0707.2474}{\tt arXiv:0707.2474}.
\bibitem[{Currie et~al.(1989)Currie, Wallace and Warren-Smith}]{sgp38}
\bibinfo{author}{Currie, M.J.}, \bibinfo{author}{Wallace, P.T.},
  \bibinfo{author}{Warren-Smith, R.F.}, \bibinfo{year}{1989}.
\newblock \bibinfo{title}{{Starlink General Paper 38}}.
\newblock \bibinfo{howpublished}{Starlink Project}.
\bibitem[{Demleitner et~al.(2014)Demleitner, Greene, Le~Sidaner and
  Plante}]{ACVOregistry}
\bibinfo{author}{Demleitner, M.}, \bibinfo{author}{Greene, G.},
  \bibinfo{author}{Le~Sidaner, P.}, \bibinfo{author}{Plante, R.L.},
  \bibinfo{year}{2014}.
\newblock \bibinfo{title}{{The virtual observatory registry}}.
\newblock \bibinfo{journal}{Astronomy and Computing}
  \DOIprefix\doi{10.1016/j.ascom.2014.07.001}.
\bibitem[{Erard et~al.(2014)Erard, Cecconi, Le~Sidaner, Berthier, Henry,
  Chauvin, Andr\'{e}, G\'{e}not, Jacquey, Gangloff, Bourrel, Schmitt, Capria
  and Chanteur}]{ACVOplanetary}
\bibinfo{author}{Erard, S.}, \bibinfo{author}{Cecconi, B.},
  \bibinfo{author}{Le~Sidaner, P.}, \bibinfo{author}{Berthier, J.},
  \bibinfo{author}{Henry, F.}, \bibinfo{author}{Chauvin, C.},
  \bibinfo{author}{Andr\'{e}, N.}, \bibinfo{author}{G\'{e}not, V.},
  \bibinfo{author}{Jacquey, C.}, \bibinfo{author}{Gangloff, M.},
  \bibinfo{author}{Bourrel, N.}, \bibinfo{author}{Schmitt, B.},
  \bibinfo{author}{Capria, M.T.}, \bibinfo{author}{Chanteur, G.},
  \bibinfo{year}{2014}.
\newblock \bibinfo{title}{Planetary science virtual observatory architecture}.
\newblock \bibinfo{journal}{Astronomy and Computing} \DOIprefix\doi{DOI:
  10.1016/j.ascom.2014.07.005}.
\bibitem[{Fitzpatrick et~al.(2013)Fitzpatrick, Laurino, Paioro and
  Taylor}]{adassxxii_bof}
\bibinfo{author}{Fitzpatrick, M.}, \bibinfo{author}{Laurino, O.},
  \bibinfo{author}{Paioro, L.}, \bibinfo{author}{Taylor, M.B.},
  \bibinfo{year}{2013}.
\newblock \bibinfo{title}{Application interoperability with samp}, in:
  \bibinfo{editor}{Friedel, D.}, \bibinfo{editor}{Freemon, M.},
  \bibinfo{editor}{Plante, R.} (Eds.), \bibinfo{booktitle}{ADASS XXII},
  \bibinfo{publisher}{ASP}, \bibinfo{address}{San Francisco}. p.
  \bibinfo{pages}{395}.
\bibitem[{G\'{e}not et~al.(2014)G\'{e}not, Andr\'{e}, Cecconi, Bouchemit,
  Budnik, Bourrel, Gangloff, Dufourg, Hess, Modolo, Renard, Lormant, Beigbeder,
  Popescu and Toniutti}]{ACVOspacephys}
\bibinfo{author}{G\'{e}not, V.}, \bibinfo{author}{Andr\'{e}, N.},
  \bibinfo{author}{Cecconi, B.}, \bibinfo{author}{Bouchemit, M.},
  \bibinfo{author}{Budnik, E.}, \bibinfo{author}{Bourrel, N.},
  \bibinfo{author}{Gangloff, M.}, \bibinfo{author}{Dufourg, N.},
  \bibinfo{author}{Hess, S.}, \bibinfo{author}{Modolo, R.},
  \bibinfo{author}{Renard, B.}, \bibinfo{author}{Lormant, N.},
  \bibinfo{author}{Beigbeder, L.}, \bibinfo{author}{Popescu, D.},
  \bibinfo{author}{Toniutti, J.P.}, \bibinfo{year}{2014}.
\newblock \bibinfo{title}{{Joining the yellow hub: Uses of the Simple
  Application Messaging Protocol in Space Physics analysis tools}}.
\newblock \bibinfo{journal}{Astronomy and Computing}
  \DOIprefix\doi{10.1016/j.ascom.2014.07.007}.
\bibitem[{Hanisch et~al.(2010)Hanisch, Arviset, Genova and Rino}]{ivoadoc}
\bibinfo{author}{Hanisch, R.J.}, \bibinfo{author}{Arviset, C.},
  \bibinfo{author}{Genova, F.}, \bibinfo{author}{Rino, B.},
  \bibinfo{year}{2010}.
\newblock \bibinfo{title}{IVOA Document Standards, Version 1.2}.
\newblock \bibinfo{type}{Technical Report}. IVOA Recommendation.
\bibitem[{{Joye} and {Mandel}(2003)}]{2003ASPC..295..489J}
\bibinfo{author}{{Joye}, W.A.}, \bibinfo{author}{{Mandel}, E.},
  \bibinfo{year}{2003}.
\newblock \bibinfo{title}{{New Features of SAOImage DS9}}, in:
  \bibinfo{editor}{{Payne}, H.E.}, \bibinfo{editor}{{Jedrzejewski}, R.I.},
  \bibinfo{editor}{{Hook}, R.N.} (Eds.), \bibinfo{booktitle}{Astronomical Data
  Analysis Software and Systems XII}, p. \bibinfo{pages}{489}.
\bibitem[{{Lafrasse} et~al.(2012){Lafrasse}, {Bourges} and
  {Mella}}]{2012ASPC..461..379L}
\bibinfo{author}{{Lafrasse}, S.}, \bibinfo{author}{{Bourges}, L.},
  \bibinfo{author}{{Mella}, G.}, \bibinfo{year}{2012}.
\newblock \bibinfo{title}{{SAMP App Launcher: An On-Demand VO Application
  Starter by JMMC}}, in: \bibinfo{editor}{{Ballester}, P.},
  \bibinfo{editor}{{Egret}, D.}, \bibinfo{editor}{{Lorente}, N.P.F.} (Eds.),
  \bibinfo{booktitle}{Astronomical Data Analysis Software and Systems XXI}, p.
  \bibinfo{pages}{379}.
\bibitem[{Laurino et~al.(2014)Laurino, Budynkiewicz, {D'Abrusco}, Bonaventura,
  Busko, Cresitello-Dittmar, Doe, Ebert, Evans, Norris, Pevunova, Refsdal,
  Thomas and Thompson}]{ACVOiris}
\bibinfo{author}{Laurino, O.}, \bibinfo{author}{Budynkiewicz, J.},
  \bibinfo{author}{{D'Abrusco}, R.}, \bibinfo{author}{Bonaventura, N.},
  \bibinfo{author}{Busko, I.}, \bibinfo{author}{Cresitello-Dittmar, M.},
  \bibinfo{author}{Doe, S.M.}, \bibinfo{author}{Ebert, R.},
  \bibinfo{author}{Evans, J.D.}, \bibinfo{author}{Norris, P.},
  \bibinfo{author}{Pevunova, O.}, \bibinfo{author}{Refsdal, B.},
  \bibinfo{author}{Thomas, B.}, \bibinfo{author}{Thompson, R.},
  \bibinfo{year}{2014}.
\newblock \bibinfo{title}{{Iris: An extensible application for building and
  analyzing spectral energy distributions}}.
\newblock \bibinfo{journal}{Astronomy and Computing}
  \DOIprefix\doi{10.1016/j.ascom.2014.07.004}.
\bibitem[{McIlroy et~al.(1978)McIlroy, Pinson and Tague}]{mcilroy1978}
\bibinfo{author}{McIlroy, M.D.}, \bibinfo{author}{Pinson, E.N.},
  \bibinfo{author}{Tague, B.A.}, \bibinfo{year}{1978}.
\newblock \bibinfo{title}{Unix time-sharing system: Foreword}.
\newblock \bibinfo{journal}{Bell System Technical Journal}
  \bibinfo{volume}{57}, \bibinfo{pages}{1899}.
\bibitem[{Sauro et~al.(2003)Sauro, Hucka, Finney, Wellock, Bolouri, Doyle and
  Kitano}]{sbw}
\bibinfo{author}{Sauro, H.M.}, \bibinfo{author}{Hucka, M.},
  \bibinfo{author}{Finney, A.}, \bibinfo{author}{Wellock, C.},
  \bibinfo{author}{Bolouri, H.}, \bibinfo{author}{Doyle, J.},
  \bibinfo{author}{Kitano, H.}, \bibinfo{year}{2003}.
\newblock \bibinfo{title}{{Next generation simulation tools: the Systems
  Biology Workbench and BioSPICE integration.}}
\newblock \bibinfo{journal}{OMICS} \bibinfo{volume}{7},
  \bibinfo{pages}{{355--72}}.
\bibitem[{{Taylor} et~al.(2007){Taylor}, {Boch}, {Comparato}, {Taylor},
  {Winstanley} and {Mann}}]{2007ASPC..376..511T}
\bibinfo{author}{{Taylor}, J.D.}, \bibinfo{author}{{Boch}, T.},
  \bibinfo{author}{{Comparato}, M.}, \bibinfo{author}{{Taylor}, M.},
  \bibinfo{author}{{Winstanley}, N.}, \bibinfo{author}{{Mann}, R.G.},
  \bibinfo{year}{2007}.
\newblock \bibinfo{title}{{Binding Applications Together with PLASTIC}}, in:
  \bibinfo{editor}{{Shaw}, R.A.}, \bibinfo{editor}{{Hill}, F.},
  \bibinfo{editor}{{Bell}, D.J.} (Eds.), \bibinfo{booktitle}{Astronomical Data
  Analysis Software and Systems XVI}, p. \bibinfo{pages}{511}.
\bibitem[{{Taylor}(2005)}]{2005ASPC..347...29T}
\bibinfo{author}{{Taylor}, M.B.}, \bibinfo{year}{2005}.
\newblock \bibinfo{title}{{TOPCAT {\&} STIL: Starlink Table/VOTable Processing
  Software}}, in: \bibinfo{editor}{{Shopbell}, P.}, \bibinfo{editor}{{Britton},
  M.}, \bibinfo{editor}{{Ebert}, R.} (Eds.), \bibinfo{booktitle}{Astronomical
  Data Analysis Software and Systems XIV}, p.~\bibinfo{pages}{29}.
\bibitem[{Taylor et~al.(2012a)Taylor, Boch, Fay, Fitzpatrick and
  Paioro}]{adassxxi_paper}
\bibinfo{author}{Taylor, M.B.}, \bibinfo{author}{Boch, T.},
  \bibinfo{author}{Fay, J.}, \bibinfo{author}{Fitzpatrick, M.},
  \bibinfo{author}{Paioro, L.}, \bibinfo{year}{2012}a.
\newblock \bibinfo{title}{Samp: Application messaging for desktop and web
  applications}, in: \bibinfo{editor}{Ballester, P.}, \bibinfo{editor}{Egret,
  D.}, \bibinfo{editor}{Lorente, N.P.F.} (Eds.), \bibinfo{booktitle}{ADASS
  XXI}, \bibinfo{publisher}{ASP}, \bibinfo{address}{San Francisco}. p.
  \bibinfo{pages}{279}.
\bibitem[{Taylor et~al.(2012b)Taylor, Boch, Fitzpatrick, Allan, Paioro, Taylor
  and Fay}]{samp_std}
\bibinfo{author}{Taylor, M.B.}, \bibinfo{author}{Boch, T.},
  \bibinfo{author}{Fitzpatrick, M.}, \bibinfo{author}{Allan, A.},
  \bibinfo{author}{Paioro, L.}, \bibinfo{author}{Taylor, J.},
  \bibinfo{author}{Fay, J.}, \bibinfo{year}{2012}b.
\newblock \bibinfo{title}{Simple Application Messaging Protocol, Version 1.3}.
\newblock \bibinfo{type}{Technical Report}. IVOA Recommendation.
\newblock \href{http://arxiv.org/abs/1110.0528}{\tt arXiv:1110.0528}.
\bibitem[{\v{S}koda et~al.(2014)\v{S}koda, Draper, Castro~Neves,
  Andre\v{s}i\v{c} and Jenness}]{ACVOsplat}
\bibinfo{author}{\v{S}koda, P.}, \bibinfo{author}{Draper, P.W.},
  \bibinfo{author}{Castro~Neves, M.}, \bibinfo{author}{Andre\v{s}i\v{c}, D.},
  \bibinfo{author}{Jenness, T.}, \bibinfo{year}{2014}.
\newblock \bibinfo{title}{{Spectroscopic analysis in the virtual observatory
  environment with SPLAT-VO}}.
\newblock \bibinfo{journal}{Astronomy and Computing}
  \DOIprefix\doi{10.1016/j.ascom.2014.06.001}.
\bibitem[{{Winstanley} et~al.(2007){Winstanley}, {Taylor}, {Taylor}, {Noddle},
  {Gonzalez-Solares} and {Lindroos}}]{2007ASPC..376..571W}
\bibinfo{author}{{Winstanley}, N.}, \bibinfo{author}{{Taylor}, J.D.},
  \bibinfo{author}{{Taylor}, M.B.}, \bibinfo{author}{{Noddle}, K.},
  \bibinfo{author}{{Gonzalez-Solares}, E.}, \bibinfo{author}{{Lindroos}, J.},
  \bibinfo{year}{2007}.
\newblock \bibinfo{title}{{Astro Runtime: An API to the Virtual Observatory}},
  in: \bibinfo{editor}{{Shaw}, R.A.}, \bibinfo{editor}{{Hill}, F.},
  \bibinfo{editor}{{Bell}, D.J.} (Eds.), \bibinfo{booktitle}{Astronomical Data
  Analysis Software and Systems XVI}, p. \bibinfo{pages}{571}.

\end{thebibliography}

\end{document}